\documentclass[12pt]{elsarticle}
\usepackage{amsmath}
\usepackage{dcolumn}
\usepackage{bm}
\usepackage{indentfirst}
\usepackage{url}
\usepackage{color}
\usepackage{multirow}
\usepackage{mathrsfs}
\usepackage{bbm}
\usepackage{euscript}
\usepackage{amssymb}
\usepackage{extarrows}
\usepackage{geometry}
\usepackage{filecontents}
\usepackage{makecell}

\usepackage{amsmath}
\usepackage{extarrows}
\usepackage{stmaryrd}
\usepackage[ruled]{algorithm2e}
\usepackage{graphicx}
\usepackage{bm}
\usepackage[all]{xy}
\usepackage{indentfirst}
\usepackage{diagbox}
\usepackage{floatrow}
  \newfloatcommand{capbtabbox}{table}[][\FBwidth]
\usepackage{bbding}
\usepackage{color}
\usepackage{appendix}
\usepackage{pifont}
\usepackage{amsmath}
\usepackage{multirow}
\usepackage{mathrsfs}
\usepackage{bbm}
\usepackage{euscript}
\usepackage{amssymb}

\usepackage{float}
\usepackage{subfigure}

\newcommand{\cE}{\mathcal{E}}
\newcommand{\cA}{\mathcal{A}}

\newcommand{\cW}{\mathcal{W}}

\newcommand{\cD}{\mathcal{D}}

\usepackage{graphicx}
\usepackage{ae,aecompl}
\usepackage{amsmath}
\usepackage{amssymb}
\usepackage{times}
\usepackage{color}
\usepackage{bbm}
\usepackage{bm}
\usepackage{xcolor}
\usepackage{soul}

\newcommand{\mean}[1]{\ensuremath{\langle #1 \rangle}}

\newcommand{\tr}{{\rm Tr}}

\newcommand{\bu}{\mathbf{u}}
\newcommand{\bv}{\mathbf{v}}
\newcommand{\hH}{\hat{H}}
\usepackage{verbatim}

\newcommand{\ket}[1]{|#1\rangle}

\newcommand{\oH}{\overline{H}}

\usepackage{makecell}
\usepackage{graphicx}

\journal{iScience}

\begin{document}
\begin{frontmatter}

\title{Quantumness Speeds up Quantum Thermodynamics Processes}

\author{Ming-Xing Luo $^{1,2,3,*}$ }
\address{$^{1}$School of Information Science and Technology, Southwest Jiaotong University, Chengdu 610031, China
\\
$^2$CAS Center for Excellence in Quantum Information and Quantum Physics, Hefei, 230026, China}
\cortext[correspondingauthor]{Correspondence: mxluo@swjtu.edu.cn (M.L.)}
\address{$^{3}$Lead contact: Further information and requests for resources should be directed to the lead contact (M.X.Luo, mxluo@swjtu.edu.cn)}

\begin{abstract}
Quantum thermodynamic process involves manipulating and controlling quantum states to extract energy or perform computational tasks with high efficiency. There is still no efficient general method to theoretically quantify the effect of the quantumness of coherence and entanglement in work extraction. In this work, we propose a thermodynamics speed to quantify the extracting work. We show that the coherence of quantum systems can speed up work extracting with respect to some cyclic evolution beyond all incoherent states. We further show the genuine entanglement of quantum systems may speed up work extracting beyond any bi-separable states. This provides a new thermodynamic method to witness entangled systems without state tomography.
\end{abstract}

\end{frontmatter}
\date{\today}


\section*{Introduction}

Quantum thermodynamics provides a bridge to explore energy transfer and conversion of two systems at the microscopic level. By incorporating quantum effects into thermodynamic systems, it can gain insights into the fundamental limits of energy extraction and charging behavior of small-scale devices$^{1-4}$. So far, quantum thermodynamics has intrigued great improvements in energy storage and transfer, such as quantum heat engines$^{5-7}$ and quantum refrigerators$^{7-9}$. Quantum thermodynamics sheds light on the fundamental principles governing the behavior of quantum systems, paving the way for quantum computing and quantum information processing$^{10,11}$.

Exploiting the quantum features of the coherence or entanglement beyond the classical counterparts is one of the most important tasks in quantum thermodynamics$^{7,12,13}$. This can trace back to a basic problem from the birth of thermodynamics in 1824, i.e., what criteria is useful to compare different states with respect to their energies. In the thermodynamic process, a quantum system can provide energy for the other systems or be charged by others. This allows building battery-like quantum devices$^{14}$. Lots of potential examples are proposed from qubits$^{15}$, spins$^{16-18}$, flywheels$^{19,20}$, to collision model$^{21-23}$. In general, it require auxiliary systems to control the batteries$^{24-28}$. Moreover, with the coherence some quantum batteries show interesting features of faster and higher-power charging capability than the classical counterparts$^{14,25-31}$. This is recently extended for entangled quantum systems$^{32-40}$. Specially, most battery systems will generate entanglement under global evolution, and then show an advantage over those with only local operations$^{32,37,41,42}$. These results show new quantum effects of the coherence and entanglement in the thermodynamic processes of quantum batteries.

Recently, most quantum work extracting protocols focus on exploiting the optimal final energy that can be exchanged for a given quantum battery under cyclic control$^{3}$. This intrigues to define the maximum of the amount of extracted work as so-called ergotropy. As for the charging process, the maximum of the amount of charged work as so-called antiergotropy. Combining both quantities allows characterizing a given quantum battery using a fundamental quantity of quantum battery capacity$^{43}$. Dividing these quantities by the evolution time period provides bare Hamiltonian-based ways to characterize the energy transferring or charging processes$^{27}$.

\begin{figure}[!]
\includegraphics[width=0.6\columnwidth]{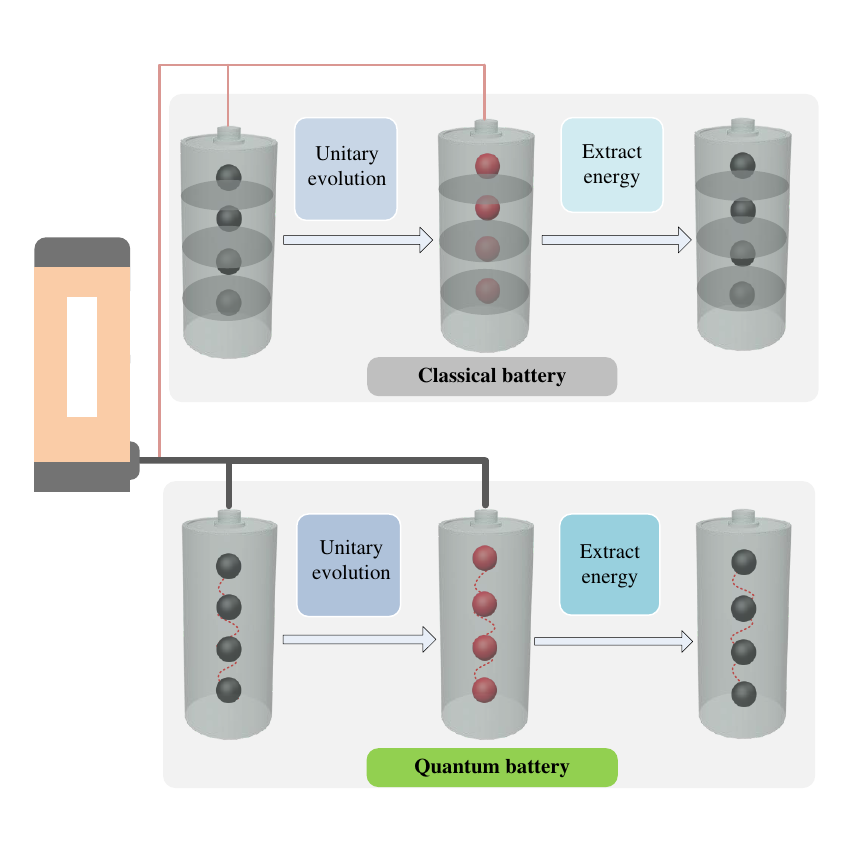}
\caption{\textbf{Schematic thermodynamic process of battery}. A classical battery system undergoes a cyclic controlling with the probing Hamiltonian within a time period in the up protocol. The quantum battery system undergoes a cyclic controlling with the same probing Hamiltonian within the same time period in the down protocol. The final extracted energy is defined according to a given bare Hamiltonian. The goal is to characterize the difference of the maximal energy-exchange speed over time-independent probing Hamiltonians.}
\label{fig-1}
\end{figure}

In this paper, we define a \emph{quantum energy-exchange speed} as a figure of the thermodynamic energy transferring that will link to quantum features of the coherence$^{44}$ and entanglement$^{47}$. We define an energy-exchange speed as the operational ratio of the Hellinger distance$^{51}$ of two final extractable energies of one battery in two complement thermodynamic procedures. This quantity is independent of the governor Hamiltonian and shows an operational relationship to the statistical speed of quantum Fisher information$^{52,53}$. We further show the present quantity is the evolution speed of energies on a unit geometric sphere. We propose a novel approach to exploit the energy extraction of coherent or entangled battery systems based on the thermodynamics speed, as shown in \cite{fig-1}.  We prove that quantum coherence can speed up the quantum thermodynamic process beyond all incoherent states. We then show the entangled battery systems can provide a speed up beyond separable states, even quadratic speeding up in an Ising model. This allows witnessing both coherence and entanglement with new thermodynamic features.



\section*{Results}

\subsection*{Quantum energy-exchange speed}

Consider a single-particle quantum $B$ which can exchange energy with external sources. Define the evolution of $B$ is generated by a Hamiltonian $\hat{H}(t)$ which accounts for energy exchanging. Assume $B$ is in a cyclic process, in which originally isolated is coupled at the time $t=0$ to external sources, and decouples from them at the time $t$ $^{2}$. The bare Hamiltonian is defined by $H_0$. The battery is thermally isolated but may involve energy exchanges between its parts,  i.e., it does not involve heat exchange with a thermal environment. The initial extractable energy is defined by
\begin{eqnarray}
E_0=\tr(\rho_0 H_0)
\end{eqnarray}
with an initial state $\rho_0$ $^{2}$. The final extractable energy is then given by $E_t=\tr(\rho(t) H_0)$, where the final state $\rho(t)$ is defined according to the Liouville-Neumann equation
\begin{eqnarray}
i\hbar\dot{\rho}=[H(t),\rho(t)].
\end{eqnarray}
For continuous systems the optimal energy depends on von Neumann entropy of final Gibbsian state $^{3}$. But this is not always right for finite systems.

In what follows, we consider a finite-dimensional single-particle battery system. More precisely, define the evolution of $B$  under a unitary operation $U$ as $\rho(t)=U\rho_0U^\dag$. The bare Hamiltonian $H_0$ is decomposed into
\begin{eqnarray}
H_0=\sum_{j=0}^{d-1}\lambda_jE|\lambda_j\rangle\langle \lambda_j|,
\end{eqnarray}
where $E$ denotes the unit energy and eigenenergies $\lambda_i$ satisfy $\lambda_0\leq \cdots \leq \lambda_{d-1}$. Without loss of generality, we associate zero energy to the lowest energetic state $\ket{\lambda_0}$. When the quantum battery goes a cyclic evolution $U_t=e^{-it \hH}$ with a time-independent Hamiltonian $\hH$, the final extractable energy is given by$^{55,56}$:
\begin{eqnarray}
E_{U_t}(\rho_0;H_0) \equiv \tr(U_t\rho_0 U_t^\dagger H_0).
\label{def_WU}
\end{eqnarray}

We extend the result to multiparticle quantum batteries. Suppose an isolated $N$-particle quantum battery is in the initial state $\rho_0$ on Hilbert space. Intead of the interaction-free Hamiltonian $\hH(t)=\sum_{i=1}^NH_{i}$ with Hamiltonian $H_{i}$ of the $i$-th particle, we consider the battery equipped with a total global Hamiltonian $\hH$. Suppose the quantum battery is subjected to a cyclic evolution with a $k$-body correlated time-independent Hamiltonian $\hH$ given by
\begin{eqnarray}
\label{H0H1}
\hH = \frac{1}{d}\sum_{i=1}^N \alpha_i\sigma_{\bv_i}^{(i)}+\frac{\gamma}{d^k} \sum_{\substack{i_1,\cdots, i_k=1}}^N\beta_{i_1\cdots i_k} \sigma_{\bu_{i_1}}^{(i_1)}\cdots \sigma_{\bu_{i_k}}^{(i_k)},
\end{eqnarray}
where $\alpha_i\in [0,1]$ account for the strength of local operations, and $\beta_{i_1\cdots{}i_k}$ denote $k$-body correlating parameters satisfying the symmetry condition of
$\beta_{i_1\cdots{}i_k}=\beta_{\tau(i_1\cdots{}i_k)}$ for any different $i_1, \cdots, i_k$ and permutation $\tau$ in the permutation group $S_k$. $\gamma$ is an arbitrary real number. Here, $\sigma^{(i_j)}_{\bu_{i_j}} \equiv \vec{\sigma}^{(i_j)}\cdot \bu_{i_j}$ are $d\times d$ generalized Pauli matrices of the $j$-th particle, where  $\vec{\sigma}$ denotes the $d$-dimensional Gell-Mann matrix vector and $\bu_{i_j}$ is a unit vector on the Gell-Mann sphere$^{54}$. After the evolution period $t$, the final state is given by $\rho=e^{-it\hH}\rho_0e^{it\hH}$. The final extractable energy has the same form (\ref{def_WU}). This intrigues us to identify the difference of extractable energies between a given evolution period. Specially, when $E_{U_0}\geq E_{U_t}$, the difference of the extractable energy gives as  $W\equiv E_{U_0}-E_{U_t}$, which characterizes the work \textit{extracted} from the battery. On the other hand, it exhibits the amount of work being \textit{charged} into the battery from other systems when $E_{U_0}\leq E_{U_t}$. The corresponding maximal works under any unitary evolutions are defined as the \emph{ergotropy} and the \emph{antiergotropy}, respectively$^{3,43}$.

Our goal in what follows is to characterize the extractable energy (\ref{def_WU}) in terms of the quantum features of battery systems. The main idea is using the thermodynamic speed to characterize the extractable energy that can be charged or extracted from the battery systems within any cyclic control protocols, as shown in Fig.1. Suppose that a given quantum battery in the state $\rho$ will be probed by applying a global transformation $U_t\equiv e^{-it\hat{H}}$ with an evolution time $t$. Define an operational measure to quantify the difference of extractable energies $E_{U_{t}}$ and $E_{U_{t'}}$ during the evolution period $[t,t']$ as
\begin{eqnarray}
 \label{Eq-Eucl}
\cD_E(t, t') &=&
\left(
\frac{1}{\tr H_0}\left(\sqrt{E_{U_t}(\rho;H_0)}-\sqrt{E_{U_{t'}}(\rho,H_0)}\right)^2
\right.
\nonumber\\
&&+
\left.\left(\sqrt{E_{U_t}(\rho;\oH_0)}-\sqrt{E_{U_{t'}}(\rho,\oH_0)}
\right)^2
\right)^{1/2},
\end{eqnarray}
where $\oH\equiv \mathbbm{1}-H_0/\tr H_0$ defines the complement Hamiltonian of $H_0$, with the corresponding final extractable energy  $E_{U_t}(\rho;\oH_0)=\tr(U_t\rho U_t^\dagger \oH_0)$. Both quantities satisfy the operational relationship of
\begin{eqnarray}
E_{U_t}(\rho,H_0)+\tr{}H_0 E_{U_t}(\rho,\oH_0)=\tr{}H_0
\end{eqnarray}
for any $U_t,\rho$ and $H_0$. The present distance (\ref{Eq-Eucl}) is zero, if and only if two final states are same. The maximal difference is $\sqrt{2}E$ if one final extractable energy is zero and the other is $E/2$. The metric (\ref{Eq-Eucl}) can be regarded as a Hellinger distance of extractable energies$^{52,53}$. This allows us to define a thermodynamic speed to figure out how fast can the extractable energy be exchanged under the cyclic control of the given battery system as:
\begin{eqnarray}
    v \equiv v(t) = \frac{d \cD_E}{d t}\Big\vert_{t},
\end{eqnarray}
i.e., the ratio at which $\cD_W$ changes with $t'$ around the reference time $t$. From the Taylor expansion it follows that
\begin{eqnarray}
v^2&=&\frac{1}{4W_{U_t}\tr{}H_0}  \dot{ W}_{U_t}^2+\frac{1}{4W_{U_t}(\rho;\oH_0)}\dot{W}_{U_t}(\rho;\oH_0)^2
\nonumber\\
&=&\frac{1}{4W_{U_t}(\tr{}H_0-W_{U_t})}\dot{W}_{U_t}^2
\label{Eq-workspeed}
\end{eqnarray}
with $\dot{ W}_{U_t}\equiv \frac{dW_{U_t}}{dt}$. The present speed will be used to explore the capability of extracting energies.

Define a maximal quantum energy-exchange speed as
\begin{eqnarray}
\label{speed}
v_w \equiv \max_{\hH}v,
\end{eqnarray}
where the maximum is over all equipped Hamiltonians in the unitary evolution protocols. This means the present quantity $v_w$ is independent of the probing Hamiltonian and provides a general feature of the controlling protocol while the known definition of quantum speed depends on the controlling Hamiltonian$^{8,27,35}$.

Inspired by the classical Fisher information$^{53}$, the present quantity (\ref{speed}) can be evaluated according to the symmetric logarithmic derivative as
\begin{eqnarray}
 v^2_w = \tr(\rho \cW^2),
 \label{spped}
\end{eqnarray}
where the symmetric logarithmic derivative $\cW$ is uniquely defined on the support of $\rho$ via the relation $\dot{\rho}= \tfrac{1}{4\sqrt{E}}(\cW \rho + \rho \cW)$. This implies the quantity $v^2_w$ is convex from the convexity of the Fisher information$^{57}$.

From Eq.(\ref{Eq-workspeed}) it follows that
\begin{eqnarray}
v_w= \frac{1}{2\sqrt{W_{U_t}(\tr{}H_0-W_{U_t})}} \dot{ W}_{U_t}
\end{eqnarray}
if $\dot{W}_{U_t}\geq 0$. By integral over time period $[0,t]$ this implies a new formula of the operational charging work as (Method):
\begin{eqnarray}
\cE(\rho,H_0)\equiv \int_{0}^t v_w dt =\arcsin\sqrt{\frac{W_{U_t}}{\tr{}H_0}}-\arcsin\sqrt{\frac{W_{U_0}}{\tr{}H_0}}
\label{Emma1}
\end{eqnarray}
under a cyclic charging evolution $U_t$. Instead, if $\dot{W}_{U_t}\leq 0$ we get a new form of the operational extracting work as
\begin{eqnarray}
\cA(\rho,H_0)=\arcsin\sqrt{\frac{W_{U_0}}{\tr{}H_0}}-\arcsin\sqrt{\frac{W_{U_t}}{\tr{}H_0}}
\label{Emmaa2}
\end{eqnarray}
under a cyclic discharging evolution $U_t$.  Both quantity $\cE$ and $\cA$ provide different metrics of quantum work procedure from previous definitions$^{43}$, as shown in Fig.2. This further intrigues to define the maximal works of \emph{ergotropy} and \emph{antiergotropy} with the present metric, which is valuable for further exploration.

\begin{figure}[!]
\includegraphics[width=0.7\columnwidth]{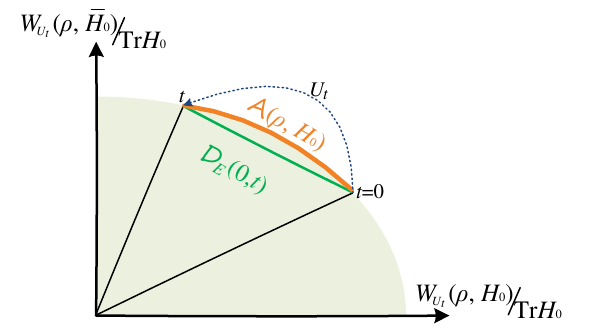}
\caption{\textbf{Operational work extracting}. For an $n$-particle quantum state $\rho$ the normalized extracting work $W_{U_t}(\rho,H_0)$ and dual work $W_{U_t}(\rho,\oH_0)$ consist of a unit cycle. When the state undergoes a unitary transformation $U_t$,  the present quantity $\cD_W$  denotes the quantum work distance (green line) while $\cA(\rho,H_0)$ denotes the evolution distance (orange arc). }
\label{fig-2}
\end{figure}

Now, we present the main result to show the energy-exchange speed of quantum battery systems in the superposition states$^{44}$ and multipartite entangled states$^{47}$.  Eq.(\ref{speed}) implies this quantity is independent of all bare Hamiltonians. Given two batteries in the states $\rho$ and $\varrho$ on the same Hilbert space, suppose both are under the same control with Hamiltonian $\hH$. From Eq.(\ref{spped}), if $v_w(\rho)^2\geq v_w(\varrho)^2$, it means the battery in the state $\rho$ has a larger maximum energy-exchange speed beyond the battery in the state $\varrho$. Informally, as for any isolated quantum battery systems, its coherence can speed up the fastest energy-exchange over all the incoherent systems. This provides a new state-independent supremacy in quantum thermodynamic over all classical counterparts. As for multiparticle quantum batteries, the optimal work extraction may not generate multipartite entanglement$^{33}$. However, we will show the genuine entanglement$^{63}$ may show a state-dependent supremacy to speed up the fastest energy-exchange over biseparable systems. Here, the genuine entanglement means it cannot be decomposed into a mixture of any biseparable state $\rho_{S_1}\otimes\rho_{S_2}$, where $S_1$ and $S_2$ denotes a bipartition of all particles.

\textbf{Theorem 1}. Given an $n$-particle quantum battery in the pure state $|\phi\rangle$, the following results hold:
\begin{itemize}
    \item[(i)]  there exists a time-independent probing Hamiltonian $\hH$ such that $v_w(\ket{\phi})^2 >v_w(\rho_{ic})^2$ for any incoherent states $\rho_{ic}$ if $|\phi\rangle$ is a coherent state;
   \item[(ii)] there exists a time-independent probing  Hamiltonian $\hH$ such that $v_w(\ket{\phi})^2 \geq v_w(\rho_{bs})^2$ for any bi-separable state $\rho_{bs}$ if $|\phi\rangle$ is genuinely entangled state.
\end{itemize}

The proof is inspired by recent methods$^{53,58}$. For any pure states and a control protocol with the time-independent probing Hamiltonian $\hH$, the quantum energy-exchange speed can be evaluated as
\begin{eqnarray}
v_w^2=E\mean{(\Delta \hH)^2}.
\label{genspeed}
\end{eqnarray}
To show the effect of quantum coherence, from the convexity of quantum energy-exchange speed, it is sufficient to prove the result for all the isolated systems, i.e., the pure states. A simple fact is that any two different incoherent states are orthogonal. This allows finding a simple Hamiltonian for each coherent state such that the average energy-exchange speed is larger than its of any incoherent states. The detailed proof is shown in Method. For multiparticle entangled batteries, the quantum energy-exchange speed depends on specific Hamiltonian $H$ and its state, but is no less than its of any bi-separable batteries.

\subsection*{Coherent quantum batteries}

Quantum coherence refers to the state of a quantum system where its constituent particles are in a superposition$^{44,45}$. Quantum coherent states are characterized by their ability to exhibit interference effects, making them valuable for applications such as quantum computing, quantum communication, and quantum metrology. We quantify the maximal energy-exchange speed of a given coherent battery. We estimate the largest energy-exchange speed over all classically correlated  states. From the convexity, we obtain the following inequality
\begin{eqnarray}
\label{Eq-statspeedich}
v^2 \leq v^2_{ic} \equiv\max_{\ket{\psi_{ic}}} v_{w}^2,
\end{eqnarray}
where the maximum is over all pure product incoherent states $\ket{\psi_{ic}} \in \{\otimes_{i=1}^N \ket{i_j}, i_j=0, \cdots, d-1\}$. Specially, for a given quantum battery the maximum energy-exchange speed can be saturated by optimal Hamiltonians$^{53}$.

As its derived in Method, we prove that the maximum energy-exchange of incoherent state $\ket{\psi_{ic}}$ under the cyclic evolution with the probing Hamiltonian (\ref{H0H1}) is given by
\begin{eqnarray}
 \label{Eq-vHich}
v_{ic}^2= \nu_{1}^2 +\nu_{2}^2\gamma+\nu_3^2\gamma^2,
\end{eqnarray}
where $\nu_1=(\Delta \hH_1)^2$, $\nu_2=\mean{\{\hH_1,\hH_2\}}$  and $\nu_3=(\Delta \hH_2)^2$ with local probing Hamiltonians $\hH_1$ and nonlocal probing Hamiltonians $\hH_2$, i.e., $\hH=\hH_1+\hH_2$. All the details of these correlations are shown in Method.  As $v_{ic}$ bounds the energy-exchange speed over all probing Hamiltonians and all incoherent states, the coherent battery system has a larger maximum energy-exchange speed than incoherent systems if the battery states violate the inequality (\ref{Eq-statspeedich}).

\textit{Example 1}. Consider a multiple-qubit battery with a constant Hamiltonian, i.e., $\gamma=0$. From Eq.(\ref{Eq-vHich}) we obtain the maximum energy-exchange speed as
\begin{eqnarray}
v_{ic}^2= \frac{E}{8}\sum_{\sigma^{(i)}_{\bv_i}\not=\sigma_z} \alpha_i^2+ \frac{E}{4}\sum_{\sigma^{(i)}_{\bv_i}=\sigma_z} \alpha_i^2.
\label{vich}
\end{eqnarray}
This implies that for a product coherent pure state $\ket{+}^{\otimes N}$, with the homogeneous Hamiltonian of $\alpha_i=a>0$, from Eq.(\ref{genspeed}) the local probing Hamiltonian $\hH=\sigma_z^{\otimes N}$ gives the maximum energy-exchange speed satisfies $v_{Q}^2=N^2a^2/4$, which provides a quadratic speed up beyond the maximum energy-exchange speed $v_{ic}^2=Na^2/4$ for all incoherent states. For general coherent batteries,  the maximum energy-exchange speed depends on nonlocal probing Hamiltonians.

\subsection*{Entangled batteries}

Entanglement is a fundamental concept in quantum mechanics$^{46-48}$. For an entangled two particles measuring one particle can instantaneously affect the state of the other$^{46}$. This phenomenon cannot be described by a classical understanding of cause and effect$^{49,50}$. Here, we quantify the maximum energy-exchange speed for an entangled battery. From the convexity of the maximum energy-exchange speed we obtain the following inequality
\begin{eqnarray}
\label{Eq-statspeedent}
v^2\leq  v^2_{fs}\equiv \max_{\ket{\psi_{s}}} v^2_{w},  \end{eqnarray}
where the maximum is over all product pure states $\ket{\psi_{fs}} \in \{\otimes_{i=1}^N \ket{\psi^{(i)}}\}$ or all bi-separable product states $\ket{\psi_{bs}} \in \{\ket{\psi}_{S_1}\ket{\psi}_{S_2}, \forall S_1\cup S_2=\{1, \cdots, N\}, S_1\cap S_2=\emptyset\}$ for witnessing entangled or genuinely entangled batteries, respectively$^{63}$. Similar to the proof for coherent systems in Method, the maximum energy-exchange speed of $\ket{\psi_{s}}$ with the probing Hamiltonian $\hH$ has the form (\ref{Eq-vHich}) with respect to separable states. The entangled battery may violate the inequality (\ref{Eq-statspeedent}).  The general case depends on the probing Hamiltonians.

\textit{Example 2}. Consider an Ising model with the $k$-paired nearest-neighbor interaction$^{59}$, i.e., $V_{ij}=\frac{1}{2k}\sum_{\forall j,0<|j-i|\leq k}\delta(j,i)$ with Dirac delta function $\delta$  and $\bv\cdot \bu=1$. $\gamma$ denotes the ferromagnetic coupling factor. For the homogeneous case of $\alpha_i=a$ and $\bv=\bu$, the optimal fully separable state for maximizing Eq.~(\ref{Eq-vHich}) is given by $\otimes_{i=1}^N \ket{\psi_i}$, where $\ket{\psi_i}$ is a superposition state in terms of two eigenstates of $\sigma_{\bu}$. We obtain the maximum energy-exchange speed from Eq.~(\ref{Eq-vHich}) satisfy  (Method):
\begin{eqnarray}
v_{fs}^2  =\frac{NE}{4}\left\{
\begin{split}
&a^2+a_0\gamma^2+O(\gamma^4),  \gamma\leq \gamma_c;
  \\
& \frac{a^2}{k}+a\gamma+(\frac{1}{k}+\frac{k}{N})\gamma^2, \gamma\geq \gamma_c;
 \end{split}
 \right.
\label{Ising1ent1a}
\end{eqnarray}
where $a_0=(8(N-k+1)ka-Nka^2+N+k^2)/(4Nk)$. The critical value $\gamma_c$ is defined with equal of two speeds in Eq.(\ref{Ising1ent1a}). Numerical evaluations are shown in Fig.3.

\begin{figure}[!]
\includegraphics[width=\columnwidth]{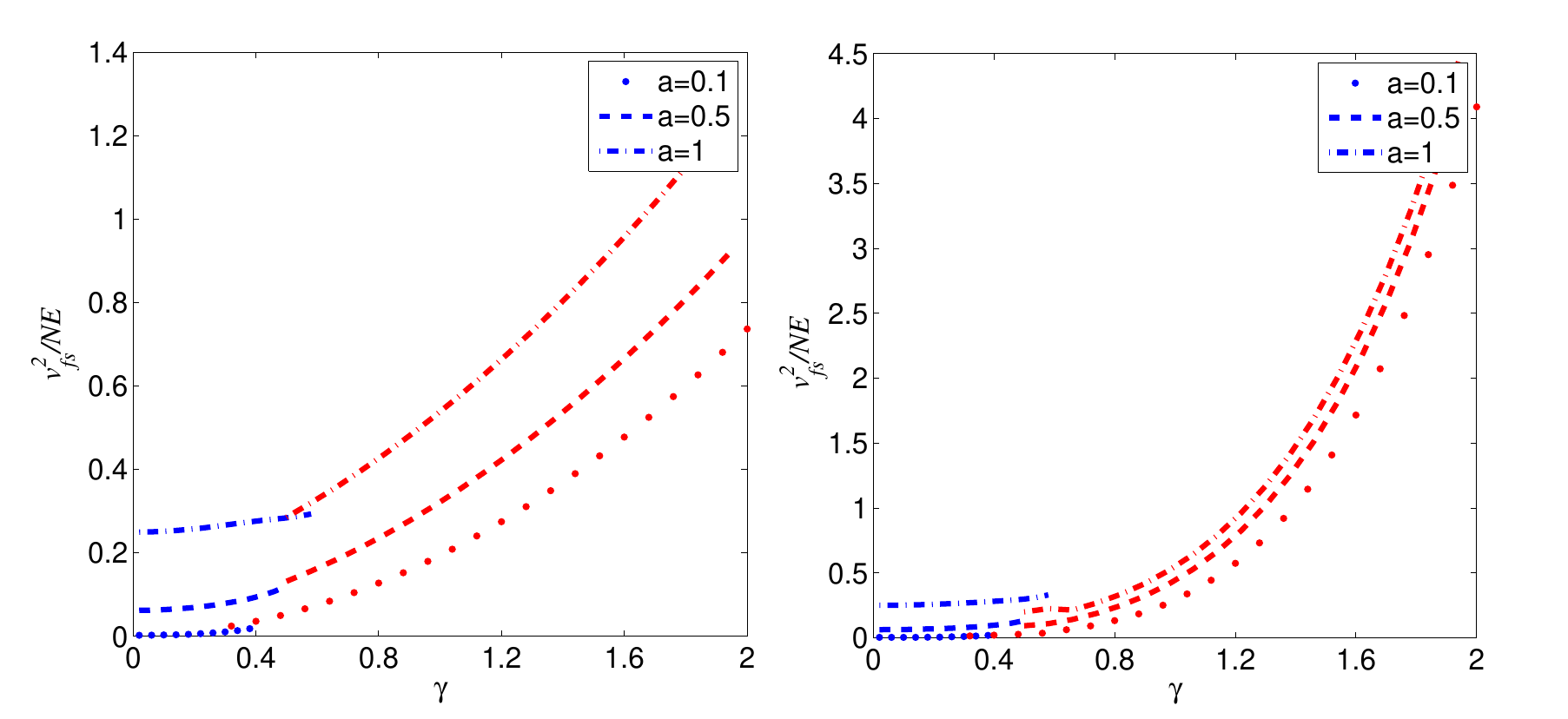}
\caption{\textbf{Witness of entangled battery with the maximum energy-exchange speed.}
Maximum energy-exchange speed of fully separable states, $v_{\max}$ (dots), probed by the Ising Hamiltonian. The present Hamiltonian $\hH$ provides a method to witness entanglement if the battery has larger energy-exchange speed than $v_{\max}$.}
\label{fig-3}
\end{figure}

Under local probing Hamiltonians, the maximum energy-exchange speed of special entangled states may be larger than its of fully separable batteries. For exploring the quantum supremacy of general entangled battery, it requires both local and nonlocal probing Hamiltonians.

\textit{Example 3}. Consider the Dicke state $|N,k\rangle$ with $k$ number of excited spins$^{60}$. If the local probing Hamiltonian $\hH_1$ satisfies $\bv\cdot\bu=0$ and $k=2$, from Eq.(\ref{genspeed}) its maximum energy-exchange speed is given by $v_Q^2=E((n+2)N-n^2)/2$. This is larger than the maximum energy-exchange speed $v_{fs}^2=EN/4$ for all fully separable states when $n<N/2$.

Consider the maximally entangled Greenberger-Horne-Zeilinger (GHZ) state$^{61}$, it is easy to show the maximum energy-exchange speed $N^2E/4$ from Eq.(\ref{genspeed}), which shows a quadratic speed up over all fully separable states with nonlocal nearest-neighbor Hamiltonians.


\section*{Discussion}

In applications, if there are some averaged Hamiltonians $\langle W \rangle_t = \sum_{i} p_i W_{i}(t)$, it is useful to extend the Hellinger distance and energy-exchange speed to the work distribution $\overline{W}(t)$, where $w_0 = \frac{1}{m} \sum_{i=1}^{m} p_iW_i$ and $W_1, \cdots, W_m$ are extracted works from different bare Hamiltonians $H_1,\cdots, H_m$. This allows to characterize the statistical mixture of quantum work speed. But we can not obtain the operational standard forms (\ref{Emma1}) and (\ref{Emmaa2}) of extracting work. This intrigues new problems to explore the relationships between the probing Hamiltonians and average energy-exchange speed.

The present energy-exchange speed reveals the quantum features of quantum batteries in a unified manner. This requires to probe a battery system with a generic multi-particle Hamiltonian. The present method shares several important properties. One is that the quantum features of the coherence and entanglement allow quantum supremacy of the maximum energy-exchange speed even quadratic speed up. This shows the importance of quantum features in quantum-size batteries for extracting or charging energies. The second is that the present quantity provides a new way to witness the quantum properties of coherence and entanglement hold in given quantum batteries. Beside the possibility to witness a larger class of entangled states, the extracting work speed allows to take in account the residual coupling among neighboring particles$^{62}$.

To conclude, the method discussed in this manuscript allows the experimental characterization of a larger class of quantum states. The present results show that entanglement can be detected even when the probing Hamiltonian is nonlinear and therefore generates entanglement. This opens the way to study entanglement near quantum phase transition points by quenching the parameters of the governing many-body Hamiltonian.

\section*{Limitations of Study}

This paper aims to show the quantum properties of coherence and entanglement can speed up quantum thermodynamics processes. The main limitation of the proposed method is from the Fish information.

\section*{Acknowledgements}

This work was supported by the National Natural Science Foundation of China (Nos. 62172341,61772437), Sichuan Natural Science Foundation (No. 2023NSFSC0447), and
Interdisciplinary Research of Southwest Jiaotong University China (No.2682022KJ004).

\section*{Author Contributions}

M.X. conceived the study and wrote the manuscript.

\section*{Declaration of Interests}

The authors declare no competing interests.

\section*{STAR*Methods}

\subsection*{Resource Availability}

\subsubsection*{Lead Contact}

Further information and requests for resources should be directed to the lead contact Ming-Xing Luo (mxluo@swjtu.edu.cn).

\subsubsection*{Materials Availability}

This study did not generate new materials.

\subsubsection*{Data and Code Availability}

This study has no data and code available.

\subsection*{Key resources table}

This study is no key resources table.

\section*{Method details}

\subsection*{Proof of Eq.(13)}

Since ${W_{U_t}}\leq {\tr H_0}$ for any state $\rho$ and Hamiltonian $H_0$, define $\sin\theta\equiv \sqrt{W_{U_t}/\tr H_0}$ and $\cos\theta\equiv \sqrt{1-W_{U_t}/\tr H_0}$. This implies that
\begin{eqnarray}
\Delta \cD_w&\equiv &\int_{0}^t v  dt
\nonumber\\
&=&\int_{0}^t\frac{1}{2\sqrt{W_{U_t}(\tr H_0-W_{U_t})}} \dot{W}_{U_t} dt
\nonumber\\
&=&\int_{\arcsin(W_{U_t}/\tr H_0)}^{\arcsin(W_{U_0}/\tr H_0)}\frac{1}{2\sin\theta \cos\theta} d\sin^2\theta
\nonumber\\
&=&\arcsin\sqrt{\frac{W_{U_
t}}{\tr H_0}}-\arcsin\sqrt{\frac{W_{U_0}}{\tr H_0}}.
\end{eqnarray}

\subsection*{Proof of Theorem 1}

The proof is inspired by recent methods$^{53,58}$. Given a quantum battery in the pure state $\ket{\phi}$, and a unitary transformation $e^{-i\hH t}$ under the time-independent probing Hamiltonian $\hH$, the energy-exchange speed is given by
\begin{eqnarray}
v_w^2(\ket{\phi})=E\Delta \hH^2.
\label{Method22}
\end{eqnarray}
We firstly prove the result for quantum batteries in the coherent states. For a given quantum battery in the coherent state $\ket{\phi}$, let the probing  Hamiltonian be $\hH=|\psi_{ic}\rangle\langle \psi_{ic}|^{\otimes n}$, where $\ket{\psi_{ic}}$ satisfies $|\langle \psi_{ic}|\phi\rangle|^2\equiv\lambda \not=1/2$. This implies from Eq.(\ref{Method22}) that
\begin{eqnarray}
v(\ket{\phi})^2&=&E\mean{\Delta \hH^2}_{\ket{\phi}}
\nonumber\\
&=&E\mean{\hH-\mean{\hH}^2}_{\ket{\phi}}
\nonumber\\
&=&E(\lambda-\lambda^2)>0.
\end{eqnarray}

But for any incoherent state $|\phi_{ic}\rangle$, we obtain that
\begin{eqnarray}
v(\ket{\phi_{ic}})^2&=&E\mean{\Delta \hH^2}_{\ket{\phi_{ic}}}
\nonumber\\
&=&E\mean{\hH-\mean{\hH}^2}_{\ket{\phi_{ic}}}
\nonumber\\
&=&0,
\end{eqnarray}
where we have used the fact $\mean{\hH}_{\ket{\phi_{ic}}}\in \{0,1\}$ for incoherent states $\ket{\phi_{ic}}$. This implies that $v(\ket{\phi})^2>\max_{\ket{\phi_{ic}}}v(\ket{\phi_{ic}})^2$.

Now, we show the result for genuinely entangled batteries. Consider a given isolated battery system in the entangled pure state $\ket{\phi}$. If one makes use of the special entanglement witness operator$^{47}$ of $\cW_{\phi}\equiv|\phi\rangle\langle \phi|-\lambda\mathbbm{1}$ as the probing Hamiltonian $\hH$, it follows that $v(\ket{\phi})^2=0$ and $v(\ket{\phi_{ic}})^2\geq 0$, where  $\lambda=\max_{\rho_{ic}}\tr(\rho_{ic}|\phi\rangle\langle \phi|)$. This means the well-known entanglement witness is useless to design the probing Hamiltonian. Instead, we use the following fact, i.e., there is a biseparable pure state $\ket{\phi_{bs}}$ such that
$|\langle\phi_{bs}|\phi\rangle|^2=1/2$. This can be proved by using Schmidt decomposition of $\ket{\phi}$ in terms of any bipartition of all particles. With this state, we define the probing Hamiltonian as $\hH=|\phi_{bs}\rangle \langle \phi_{bs}|$. It follows that
\begin{eqnarray}
v(\ket{\phi})^2&=&E\mean{\Delta \hH^2}_{\ket{\phi}}
\nonumber\\
&=&E\mean{\hH-\mean{\hH}^2}_{\ket{\phi}}=\frac{E}{4}.
\end{eqnarray}

For any biseperable pure state $\ket{\psi_{bs}}$, we obtain that
\begin{eqnarray}
v(\ket{\psi_{bs}})^2&=&E\mean{\Delta \hH^2}_{\ket{\psi_{bs}}}
\nonumber\\
&=&E(\mean{\hH}_{\ket{\psi_{bs}}}-\mean{\hH}^2_{\ket{\psi_{bs}}})\leq \frac{E}{4}.
\end{eqnarray}
This has proved the result for genuinely entangled batteries.

Similar proof holds for entangled batteries by ruling out all fully separable batteries if there is a fully separable state $\ket{\phi_{fs}}$ such that
$|\langle\phi_{fs}|\phi\rangle|^2=1/2$.

\subsection*{Proof of Eq.~(\ref{Eq-vHich})}

In this section, we evaluate the energy-exchange speed of incoherent systems as
\begin{eqnarray}
 \label{C-3}
v_{ic}^2= \nu_{1}^2 +\nu_{2}^2\gamma+\nu_3^2\gamma^2,
\end{eqnarray}
where $\nu_i$ are given by
\begin{eqnarray*}
\nu_1^2 &=&\frac{E}{d^2}\sum_{i=1}^N \alpha_i^2(1-s_{i}^2),
\\
\nu_2^2&=&\frac{4E}{d^{k+1}} \sum_{\bf j} \alpha_{i}\beta_{\bf j} (\bv_{j_t}^{(j_t)}\cdot{} \bu_{j_t}^{(j_t)}
-(s_{j_t}^{(j_t)})^2) \prod_{\ell\not=t}s_{\bu_{j_\ell}}^{(j_\ell)},
\\
\nu_3^2
&=& \frac{E}{d^{2k}} \left[\sum_{\bf i} \beta^2_{\bf i}(1-(s_{\bu_{i_1}}^{(i_1)}\cdots s_{\bu_{i_k}}^{(i_k)})^2)
+\sum_{{\bf i,j}, i_t\neq j_t, \exists t}
\beta_{\bf i} \beta_{\bf j}(1-
\prod_{i_t=j_\ell}(s_{\bu_{i_t}}^{(i_t)})^2)\prod_{i_t\not=j_t}s_{\bu_{i_t}}^{(i_t)}\right]
\end{eqnarray*}
with $s_{i}=\mean{\sigma_{\bv_i}^{(i)}}$, $s_{\bu_{i}}^{(i)}=\mean{\sigma_{\bu_i}^{(i)}}$, $\mathbf{i}=i_1\cdots{}i_k$ and $\mathbf{j}=j_1\cdots{}j_k$.

Inspired by recent methods$^{53,58}$, for pure states and unitary transformation $e^{-i\hH t}$, the energy-exchange speed is given by $v_w^2(\gamma) =E\Delta \hH^2$. Taking $\hH = \hH_1 + \gamma\hH_2$, this equality implies that
\begin{eqnarray}
\nu_1^2 &=& E\Delta \hH_1^2,
\label{v1ich}
\\
\nu_2^2 &=&E(\mean{ \{\hH_1, \hH_2\}} - 2 \mean{\hH_1} \mean{\hH_2}),
\label{v3ich}
\\
\nu_3^2 &=& E\Delta \hH_2^2.
\label{v2ich}
\end{eqnarray}
From the convexity of the energy-exchange speed, it is sufficient to show $v_{\max}$ for all product incoherent pure states.

We first evaluate $\nu_1^2$ with respect to product incoherent pure states. For the single-particle Hamiltonian $\hH_1$ given by $\hH_1=\frac{1}{d}\sum_{i=1}^N \alpha_i\sigma_{\bv_i}^{(i)}$, we obtain that
\begin{eqnarray}
\mean{\hH_1^2}&=&
\frac{1}{d^2}\sum_{i,j=1}^N \alpha_i\alpha_j\mean{\sigma_{\bv_i}^{(i)}\sigma_{\bv_j}^{(j)}}
 \nonumber\\
&:=&
\frac{1}{d^2}\sum_{i,j=1}^N \alpha_i\alpha_js_{ij},
\end{eqnarray}
where $s_{ij}=\mean{\sigma_{\bv_i}^{(i)}\sigma_{\bv_j}^{(j)}}$. If all vectors $\bv_j$ are the same, it follows a special case of $s_{ij}=1$ for $i=j$; and $0$ for other cases. Moreover, we obtain that
\begin{eqnarray}
\mean{\hH_1}^2&=&
\frac{1}{d^2}\sum_{i,j=1}^N \alpha_i\alpha_j\mean{\sigma_{\bv_i}^{(i)}}\mean{\sigma_{\bv_j}^{(j)}}
\nonumber\\
&:=&
\frac{1}{d^2}\sum_{i,j=1}^N \alpha_i\alpha_js_{i}s_{j},
\end{eqnarray}
where $s_{i}=\mean{\sigma_{\bv_i}^{(i)}}$. This implies from the equality (\ref{v1ich}) that
\begin{eqnarray}
\nu_1^2 &=& E(\mean{\hH_1^2}-\mean{\hH_1}^2)
 \nonumber\\
&=& \frac{E}{d^2}\sum_{i,j=1}^N \alpha_i\alpha_j (s_{ij}-s_{i}s_{j})
 \nonumber\\
&=& \frac{E}{d^2}\sum_{i=1}^N \alpha_i^2(1-s_{i}^2)
\end{eqnarray}
as $s_{ij}=s_{i}s_{j}$ for product incherent pure states.

Now, we evaluate $\nu_2^2$ with respect to product incoherent pure states. Let $\hH_2$ be given by $\hH_2=\frac{1}{d^k} \sum_{\substack{i_1,\cdots, i_k=1}}^N \beta_{i_1\cdots i_k} \sigma_{\bu_{i_1}}^{(i_1)}\cdots \sigma_{\bu_{i_k}}^{(i_k)}$. We obtain that
\begin{eqnarray}
\mean{\{\hH_1,\hH_2\}}&=& \mean{\hH_1\hH_2}+\mean{\hH_2\hH_1}
\nonumber
\\
&=&
\frac{1}{d^{k+1}}
\left[ \sum_{i,{\bf j}} \alpha_{i}\beta_{\bf j} \mean{\sigma_{\bv_{i}}^{(i)}\sigma_{\bu_{j_1}}^{(j_1)} \cdots \sigma_{\bu_{j_k}}^{(j_k)}}
+
\sum_{i,{\bf j}} \alpha_{i}\beta_{\bf j} \mean{\sigma_{\bu_{j_1}}^{(j_1)} \cdots \sigma_{\bu_{j_k}}^{(j_k)}\sigma_{\bv_{i}}^{(i)}}
\right]
\nonumber
\\
&=&
\frac{2}{d^{k+1}}
\left[ \sum_{\bf j}\alpha_{i}\beta_{\bf j} (\bv_{j_t}^{(j_t)}\cdot{} \bu_{j_t}^{(j_t)} \prod_{\ell\not=t}
s_{\bu_{j_\ell}}^{(j_\ell)}
+\sum_{j_1,\cdots, j_{k+1}} \alpha_{j_{k+1}}\beta_{\bf j}s_{\bu_{j_1}}^{(j_1)}\cdots s_{\bu_{j_{k+1}}}^{(j_{k+1})}
\right]
\end{eqnarray}
where $s_{\bu_{i}}^{(i)}=\mean{\sigma_{\bu_i}^{(i)}}$. Moreover, we obtain that
\begin{eqnarray}
\mean{\hH_1}\mean{\hH_2}&=& \mean{\hH_1\hH_2}+\mean{\hH_2\hH_1}
\nonumber
\\
&=&
\frac{1}{d^{k+1}}\sum_{i,{\bf j}} \alpha_{i}\beta_{\bf j} \mean{\sigma_{\bv_{i}}^{(i)}}\mean{\sigma_{\bu_{j_1}}^{(j_1)}} \cdots \mean{\sigma_{\bu_{j_k}}^{(j_k)}}
\nonumber
\\
&=&
\frac{1}{d^{k+1}}
\left[ \sum_{\bf j}\alpha_{i}\beta_{\bf j} (s_{j_t}^{(j_t)}s_{j_t}^{(j_t)} \prod_{\ell\not=t}
s_{\bu_{j_\ell}}^{(j_\ell)}
+\sum_{j_1,\cdots, j_{k+1}} \alpha_{j_{k+1}}\beta_{\bf j}s_{\bu_{j_1}}^{(j_1)}\cdots s_{\bu_{j_{k+1}}}^{(j_{k+1})}
\right]
\end{eqnarray}
This implies from the equality (\ref{v1ich}) that
\begin{eqnarray}
\nu_2^2 &=& E(\mean{ \{ \hH_1, \hH_2\}} - 2 \mean{\hH_1} \mean{\hH_2})
\nonumber
\\
&=&
\frac{4E}{d^{k+1}} \sum_{\bf j} \alpha_{i}\beta_{\bf j} (\bv_{j_t}^{(j_t)}\cdot{} \bu_{j_t}^{(j_t)}
-s_{j_t}^{(j_t)}s_{j_t}^{(j_t)}) \prod_{\ell\not=t}s_{\bu_{j_\ell}}^{(j_\ell)}.
\end{eqnarray}

Now, we evaluate $\nu_3^2$ with respect to product incherent pure states. We obtain that
\begin{eqnarray}
\mean{\hH_2^2}&=&
\frac{1}{d^{2k}} \sum_{\bf i,j } \beta_{\bf i} \beta_{\bf j} \mean{\sigma_{\bu_{i_1}}^{(i_1)}\sigma_{\bu_{j_1}}^{(j_1)} \cdots \sigma_{\bu_{i_k}}^{(i_k)}\sigma_{\bu_{j_k}}^{(j_k)}},
\end{eqnarray}
where ${\bf i}=i_1\cdots i_k$. Moreover, we obtain that
\begin{eqnarray}
\mean{\hH_2}^2=
\frac{1}{d^{2k}} \sum_{\bf i,j } \beta_{\bf i} \beta_{\bf j} \mean{\sigma_{\bu_{i_1}}^{(i_1)}}\mean{\sigma_{\bu_{j_1}}^{(j_1)}} \cdots \mean{\sigma_{\bu_{i_k}}^{(i_k)}}\mean{\sigma_{\bu_{j_k}}^{(j_k)}}.
\end{eqnarray}
 This implies from the equality (\ref{v2ich}) that
\begin{eqnarray}
 \nu_3^2 &=& E(\mean{\hH_2^2}-\mean{\hH_2}^2)
 \nonumber\\
&=& \frac{E}{d^{2k}} \sum_{\bf i,j }  \beta_{\bf i} \beta_{\bf j} (\mean{\sigma_{\bu_{i_1}}^{(i_1)}\sigma_{\bu_{j_1}}^{(j_1)} \cdots \sigma_{\bu_{i_k}}^{(i_k)}\sigma_{\bu_{j_k}}^{(j_k)}}
-\mean{\sigma_{\bu_{i_1}}^{(i_1)}}\mean{\sigma_{\bu_{j_1}}^{(j_1)}} \cdots \mean{\sigma_{\bu_{i_k}}^{(i_k)}}\mean{\sigma_{\bu_{j_k}}^{(j_k)}}).
\end{eqnarray}

Given a product incoherent state, for $i_t\not=j_t$ with any $t=1, \cdots, k$, we obtain that
\begin{eqnarray}
\mean{\sigma_{\bu_{i_1}}^{(i_1)}\sigma_{\bu_{j_1}}^{(j_1)} \cdots \sigma_{\bu_{i_k}}^{(i_k)}\sigma_{\bu_{j_k}}^{(j_k)}}
 =\mean{\sigma_{\bu_{i_1}}^{(i_1)}}\mean{\sigma_{\bu_{j_1}}^{(j_1)}} \cdots \mean{\sigma_{\bu_{i_k}}^{(i_k)}}\mean{\sigma_{\bu_{j_k}}^{(j_k)}}.
\end{eqnarray}
If ${\bf i}={\bf j}$, we obtain that
 \begin{eqnarray}
\mean{\sigma_{\bu_{i_1}}^{(i_1)}\sigma_{\bu_{j_1}}^{(j_1)} \cdots \sigma_{\bu_{i_k}}^{(i_k)}\sigma_{\bu_{j_k}}^{(j_k)}}=1.
 \end{eqnarray}
This implies that
\begin{eqnarray}
\nu_3^2 &=& E(\mean{\hH_2^2}-\mean{\hH_2}^2)
 \nonumber\\
&=& \frac{E}{d^{2k}} \sum_{\bf i} \beta^2_{\bf i}(1-(s_{\bu_{i_1}}^{(i_1)}\cdots s_{\bu_{j_k}}^{(j_k)})^2)
+\sum_{\substack{\bf i,j\\ i_t\neq j_t, \exists t}}
\beta_{\bf i} \beta_{\bf j}(1-
\prod_{i_t=j_\ell}(s_{\bu_{i_t}}^{(i_t)})^2)\prod_{i_x\not=b_y}s_{\bu_{i_t}}^{(i_t)}.
\end{eqnarray}

\subsection*{Ising model}

Let us consider an Ising model with the $k$-pair nearest-neighbor interaction, i.e., $V_{ij}=\frac{1}{2k}\sum_{\forall j,0<|j-i|\leq k}\delta(j,i)$ with Dirac delta function $\delta$ and $\bv\cdot \bu=1$.  For $\gamma$ smaller than a critical value $\gamma_c$, the maximum speed $v_{fs}$ is achieved when  $\mean{\sigma^{(i)}_{\bu}}$ are all equal to $s$. From Eq.~(\ref{Eq-vHich}), with forward tedious evaluations  the energy-exchange speed for fully separable states is given by
\begin{eqnarray}
\label{Ising1}
v_{fs}^2=\nu_1^2+\nu_2^2\gamma+\nu_3^2\gamma^2,
\label{Ising1sep}
\end{eqnarray}
where $\nu_i$ are given by
\begin{eqnarray*}
\nu_1^2&=&\frac{NE}{4}a^2(1-s^2),
\\
\nu_2^2&=&\frac{E}{2}a(N-k+1)(s-s^3),
\\
\nu_3^2&=&\frac{E}{4}[\frac{1}{4k}(N+k^2)+((1-\frac{1}{2k})N-\frac{5k}{6}-1)s^2
-((1-\frac{1}{4k})N-\frac{7k}{12}-\frac{9}{8})s^4].
\end{eqnarray*}
For a large $N$, $\gamma\ll 1$ and $0\leq a\leq 1$, we find $s =\gamma+O(\gamma^3)$ for optimizing $v_{fs}^2$ as
\begin{eqnarray}
 v_{fs}^2
 &=&\frac{NE}{4}\left[a^2(1-s^2)+2a(s-s^3) (1-\frac{k-1}{N})\gamma +g(s)\gamma^2\right]
 \nonumber\\
&=&\frac{NE}{4}[a^2+a_0\gamma^2+O(\gamma^4)],
\label{Ising1ent1}
\end{eqnarray}
where $a_0=(8(N-k+1)ka-Nka^2+N+k^2)/(4Nk)$.

For $\gamma>\gamma_c$, we obtain the optimized energy-exchange speed when $\mean{\sigma_{\bv}^{(i)}}=0$ for all $i$, and $\mean{\sigma_{\bu}^{(kj+1)}}=1$ and $\mean{\sigma_{\bu}^{(kj)}}=\cdots =\mean{\sigma_{\bu}^{(kj+k-1)}}=0$ for all $j$. This implies that
$\nu_1^2=Na^2E/4k, \nu_2^2=(N-1)aE/4$ and $\nu_3^2=(N+k^2)E/4k$, i.e.,
\begin{equation}
\label{Ising2ent}
v_{fs}^2(\beta) =\frac{NE}{4}
\left[ \frac{a^2}{k}+a\gamma+(\frac{1}{k}+\frac{k}{N})\gamma^2
\right].
\end{equation}
Indeed, in the limit $\gamma \gg 1$, Eq.~(\ref{Ising2ent}) goes as $NE\gamma^2/4k$ and thus is larger than Eq.~(\ref{Ising1sep}) for any $k$, which goes as $Nk(4k-1+4N)\gamma^2E/(16N(4k-1))$. The critical value of $\gamma$ is obtained setting Eq.~(\ref{Ising1sep}) is equal to the maximum over $s$ of Eq.~(\ref{Ising2ent}), and given by
\begin{eqnarray}
\gamma_c=-\frac{c_2}{2c_1}+\frac{1}{2c_1}\sqrt{c_2^2-4c_1c_3},
\end{eqnarray}
where $c_i$ are defined by
\begin{eqnarray*}
 c_1&=&
-\frac{3}{4k}-\frac{3k}{4N}+\frac{1}{N}((1-\frac{1}{2k})N-\frac{5k}{6}-1)s^2
\\
&&-\frac{1}{N}((1-\frac{1}{4k})N-\frac{7k}{12}-\frac{9}{8})s^4-\frac{1}{k}-\frac{k}{N},
\\
c_2&=&a(2(s-s^3)(N-k+1)-1),
\\
c_3&=&a^2(\frac{k-1}{k}-s^2).
\end{eqnarray*}

The upped bound of $v^2_{\max}$ for the inhomogeneous case ($\alpha_i\neq 0$) can be obtained by maximizing each term in Eq.~(\ref{Eq-vHich}) separately. This gives
\begin{eqnarray}
 \label{Eq-boundNN}
v^2_{\max}\leq
\frac{E}{4}\sum_{i=1}^N \alpha_i^2 + \frac{E\alpha_0}{4 }\gamma  + \frac{NE}{8}\gamma^2
 \end{eqnarray}
with $\alpha_0=\max\{\sum_{{\rm odd} \, i} \alpha_i, \sum_{{\rm even} \, i} \alpha_i \}$.

\end{document}